Short communication

# Determinants of gait stability while walking on a treadmill: a machine learning approach

Fabienne Reynard[a], PT MSc
Philippe Terrier[ab§], PhD

[a] Clinique romande de réadaptation SUVACare, Sion, Switzerland
[b] IRR, Institute for Research in Rehabilitation, Sion, Switzerland

**§ Corresponding author:**
Dr. Philippe Terrier
Clinique romande de réadaptation SUVACare
Av. Gd-Champsec 90
1951 Sion
Switzerland
Tel.: +41-27-603-23-91
E-mail: Philippe.Terrier@crr-suva.ch

**Abstract**

Dynamic balance in human locomotion can be assessed through the local dynamic stability (LDS) method. Whereas gait LDS has been used successfully in many settings and applications, little is known about its sensitivity to individual characteristics of healthy adults. Therefore, we reanalyzed a large dataset of accelerometric data measured for 100 healthy adults from 20 to 70 years of age performing 10 min. treadmill walking. We sought to assess the extent to which the variations of age, body mass and height, sex, and preferred walking speed (PWS) could influence gait LDS. The random forest (RF) and multiple adaptive regression splines (MARS) algorithms were selected for their good bias-variance tradeoff and their capabilities to handle nonlinear associations. First, through variable importance measure (VIM), we used RF to evaluate which individual characteristics had the highest influence on gait LDS. Second, we used MARS to detect potential interactions among individual characteristics that may influence LDS. The VIM and MARS results indicated that PWS and age correlated with LDS, whereas no associations were found for sex, body height, and body mass. Further, the MARS model detected an age by PWS interaction: on one hand, at high PWS, gait stability is constant across age while, on the other hand, at low PWS, gait instability increases substantially with age. We conclude that it is advisable to consider the participants' age as well as their PWS to avoid potential biases in evaluating dynamic balance through LDS.

**Keywords: dynamic balance; human locomotion; maximum Lyapunov exponent; nonlinear analysis;**

# 1. Introduction

Dynamic balance in human locomotion can be assessed through the local dynamic stability (LDS) method. Inspired by the maximal Lyapunov exponent that can detect deterministic chaos in nonlinear dynamic systems, LDS is a nonlinear method designed specifically to detect gait instabilities (Dingwell, 2006; Kurz et al., 2010; Terrier and Dériaz, 2013). Using LDS to characterize dynamic balance has recently become increasingly widespread. Besides several studies that highlighted the association between gait LDS and falls in elderly (Lockhart and Liu, 2008; Toebes et al., 2012), most recent examples of LDS use include: the effect of active arm swing while walking (Wu et al., 2016); the characterization of ataxic gait (Chini et al.); the effect of physical exhaustion (Hamacher et al., 2016); or the influence of inclined surfaces (Vieira et al., 2017).

Potential confounders that might influence LDS have not yet been investigated thoroughly (Bizovska et al., 2015). Several studies have focused on how walking at different speeds modifies LDS (Bruijn et al., 2009; Kang and Dingwell, 2008). However, further studies are needed to characterize the association between LDS and preferred walking speed (PWS). Several have established that LDS diminishes in older adults (Bruijn et al., 2014; Kang and Dingwell, 2008). We also highlighted an age effect among young and middle-aged adults (Terrier and Reynard, 2015). However, the effects of anthropometric characteristics (mass, height) and sex are largely unknown. Furthermore, information is lacking about how those different individual characteristics may interact to modify LDS.

Therefore, the aim of this short communication was to give further insight into the factors that may influence the dynamic balance –assessed through LDS— in a population of healthy adults. The variables of interest were: age, sex, body height and mass, and PWS. We used two complementary machine learning algorithms to evaluate the variables' ability to predict gait LDS: random forest (RF) and multivariate adaptive regression splines (MARS).

# 2. Methods

We retrospectively analyzed a dataset that was described in previous publications (Reynard and Terrier, 2014; Reynard and Terrier, 2015; Terrier and Reynard, 2015), which include details on the experimental procedure, the ethical considerations, the material, and the analytical methods. We briefly summarize hereafter the essential information.

*2.1 Subjects and experimental procedure*

One hundred healthy adults participated in the study; they were recruited across five classes of age with an equal number of males and females. The means and the standard deviations (SD) of individual characteristics across the age classes were reported in Table 1 of (Terrier and Reynard, 2015), which also shows that body mass and height, as well as PWS, did not depend on age. The participants walked 2 x 5 min. on a treadmill at PWS with a nine- day interval. PWS was measured in two steps by 1) progressively increasing the treadmill speed from a low speed and 2) progressively decreasing the treadmill speed from a high speed, until the subject reported a comfortable speed. The PWS was defined as the average speed of both tests. An inertial sensor measured trunk accelerations. Following insight gained from our previous studies (Reynard and Terrier, 2014; Reynard et al., 2014), we only analyzed the mediolateral acceleration.

*2.2 Data analysis*

The five-minute acceleration signals were truncated at 210 gait cycles (420 steps) and then resampled to a uniform length of 14,000 samples. We used Rosenstein's algorithm to assess the divergence in the state space that was reconstructed using the embedding principles; see (Terrier and Dériaz, 2013) for a presentation of the theoretical background. We used an average mutual information (AMI) algorithm to determine the average time delay (six samples). We used a global false nearest neighbors (GFNN) method to assess the dimension (six). We computed the divergence exponent over a number of samples corresponding to one step (i.e., 34). Both sessions' results were averaged. Note that a higher divergence rate indicates lower stability.

*2.3 Descriptive statistics*

A boxplot (median, quartiles and data extent) was used to show the LDS distribution among participants (Fig. 1). Four scatterplots with separated marks for males and females show bivariate distributions of LDS against other variables (Fig. 2).

*2.4 Machine learning*

As the first machine learning approach, we built an RF model with individual characteristics as predictors of LDS. In short, RF aggregates many decision trees that are built through recursive partitioning across predictors (Breiman, 2001). As the main advantage, the RF algorithm offers a variable importance measure (VIM), which can differentiate among crucial and negligible predictors (Strobl et al., 2007). In addition, RF is a robust nonparametric method that detects nonlinear associations and while considering interactions among predictors. RF is not subjected to overfitting because of its built-in resampling algorithm. Specifically, we sought 1) to assess the strength of the association between LDS and individual characteristics through RF's cross-validation capability, and 2) to classify the predictors through VIM. We used the RF implementation provided in the *R* (v. 3.3.3) package *Party* (v. 1.2-2) (Strobl et al., 2009a), which is based on conditional inference trees (Hothorn et al., 2006) and is robust to biases that may affect other RF implementations (Strobl et al., 2007). For VIM computation, we used the conditional variable importance method (Strobl et al., 2008), which does not overstate correlated predictors' importance and, hence, can be interpreted similarly to the coefficients of parametric regression models (Boulesteix et al., 2014; Strobl et al., 2009b). The procedure was as follows: 1) we declared PWS, age, body mass, and body height as continuous predictors whereas sex was encoded as a two-level categorical variable; we defined the outcome (LDS) as a continuous variable. 2) We set up the *cforest* function to grow 2,000 regression trees ('*ntree* = 2000'), each using a subset of two randomly-selected predictors for splitting ('*mtry* = 2'). We set the other tuning parameters to their default values. 3) As cross-validation based on out-of-bag (OOB) data, we used $R^2$ to assess how well the model prediction fit with the measured outcome. 4) We used the *varimp* function to evaluate the variable importance of the five predictors through the conditional importance method ('*conditional* = TRUE') (Strobl et al., 2008). Figure 2 shows the 'mean decrease in accuracy' importance scores of the predictors.

As the second machine learning approach, we selected the MARS algorithm (Friedman, 1991) because of its appropriateness to handle nonlinear relationships, its ability to detect interactions among predictors, and its good bias-variance tradeoff. In short, the MARS algorithm combines the logic of stepwise multiple regression models and regression trees. MARS fits a model with the help of piecewise linear splines as basis functions, which split predictors around knots. As in recursive partitioning trees, the MARS model is built through forward and backward phases. The forward pass recursively adds basis functions that reduce the residual error. The backward phase prunes the model to lower overfitting by penalizing model complexity.

MARS was implemented through the MATLAB (R2015a) package known as ARESLab (v. 1.13.0) (Jekabsons, 2016). We used the *aresbuild* function for piecewise-linear modelling ('cubic', false). A preliminary cross-validation analysis (command *arescvc*) had indicated that a high penalty ('$c$', 5) of generalized cross-validation (GCV) per knot decreased overall error. We also tuned the model to include potential pairwise interactions ('*maxInteraction*', 2). Finally, we set a minimal span of 15 near edges ('*useEndSpan*', 15) to lower the risk of spurious fitting at the end of data intervals. We set the other tuning parameters to their default values. The detailed equation of the MARS model is reported in Figure 3. Along with the equation, we placed fitting curves onto the scatter plots (Fig. 2), one for each continuous predictor retained by the MARS model. More precisely, the model was fed with each predictor with other predictors held constant (median). Furthermore, fitting curves were computed separately for males and females to show potential sex effects. Note that, whereas this method of representation would highlight the main effects and interactions of the sex variable, other interactions might be poorly represented. Therefore, we used 3D representation to better illustrate the interaction between two predictors (Fig. 4).

## 3. Results

The LDS distribution among participants (Fig. 1) highlights a non-skewed distribution and the presence of an outlier that was removed from subsequent analyses (final sample size, N = 99). The LDS average was 0.94 (SD = 0.07). The accuracy of the RF model based on cross-validation using OOB data is $R^2 = 0.26$. Regarding VIM, PWS and age are particularly noteworthy (Fig. 3). The MARS result (Fig. 2) indicates an age by PWS interaction. In addition, other variables were not included in the final model, which confirms the VIM analysis. Fig. 4 illustrates how the age by PWS interaction operates: older individuals with low PWS are likely to exhibit higher gait instability. On the other hand, the influence of age vanishes for individuals with a PWS higher than the average (>1.1 m/s).

## 4. Discussion

We analyzed a large dataset including 84,000 steps measured in 100 healthy adults. The results reveal that PWS and age had a relevant effect on gait stability. On the contrary, sex, mass, and height had no effect.

We are confident about the results' generalizability because of the large accelerometric dataset obtained in controlled conditions and because machine learning algorithms are robust regarding overfitting biases. In contrast, multiple regressions are prone to overfitting when a high number of predictors and interaction terms are used (Babyak, 2004). Stepwise regressions also have limitations (Thompson, 1995). In a supplementary file, we present regression analyses for comparison purpose. Note that the stepwise regression also selects age and PWS as LDS predictors.

Some limitations to our study must be considered. Mainly, this is a treadmill experiment, whose results might not be fully applicable to overground walking. Moreover, the measurement with a chest-worn accelerometer, as well as the computation of LDS, may be not fully comparable to other methods (Stenum et al., 2014).

Convergent evidence supports the hypothesis that adults in their 40s or 50s are subject to a higher fall risk while walking than are younger adults. A large retrospective study showed that middle-aged adults experienced more falls than younger adults, especially during ambulation. (Talbot et al., 2005). A cohort study analyzing the circumstances leading to a spinal cord injury showed that middle-aged patients exhibited a higher rate of injury as a result of tripping than did younger patients (Chamberlain et al., 2015). The present study's results confirm that gait becomes more unstable with age, as we already showed through classical regression analyses (Terrier and Reynard, 2015). Indeed, both RF and MARS analyses detected age as LDS predictor. MARS shows an inflexion point at 29 yr. The muscle loss that occurs at a higher pace from 40 yr. of age (Evans and Lexell, 1995) may be a factor implied in the increasing gait instability.

VIM results (Fig. 3) evidence that PWS is the most important predictor of LDS. Furthermore, MARS results (Fig. 3 & 4) unveil an age by PWS interaction: on one hand, at high PWS, gait stability is constant across age; on the other hand, at low PWS, gait instability increases substantially with age. One hypothesis could be that physical fitness–and, hence, muscle strength—was higher in a sub-sample of our participants, which allowed them to maintain a high PWS counteracting the aging effects.

## 5. Conclusions

In planning experiments for evaluating gait stability through LDS, it is advisable to consider participants' age as well as their PWS to avoid potential biases. Further studies are needed to assess relationships between gait instability and physical fitness as a potential explanation of the age by PWS interaction.

**Acknowledgments**

The study was supported by the SUVA and the Clinique Romande de Réadaptation. The Institute for Research in Rehabilitation is funded by the State of Valais and the City of Sion. Study sponsors were not implied in the study design; in the collection, analysis and interpretation of data; in the writing of the manuscript; or in the decision to submit the manuscript for publication.

**Conflict of interest statement**

There are no known conflicts of interest.

**References**

Babyak, M., 2004. What you see may not be what you get: a brief, nontechnical introduction to overfitting in regression-type models. Psychosom. Med. 66, 411–421.
Bizovska, L., Svoboda, Z., Janura, M., 2015. The possibilities for dynamic stability assessment during gait: A review of the literature. J. Phys. Edu. Sport 15, 490-497.
Boulesteix, A.-L., Janitza, S., Hapfelmeier, A., Van Steen, K., Strobl, C., 2014. Letter to the Editor: On the term 'interaction'and related phrases in the literature on Random Forests. Brief. Bioinform. 16, 338–345.
Breiman, L., 2001. Random forests. Mach. Learn. 45, 5–32.
Bruijn, S.M., van Dieen, J.H., Meijer, O.G., Beek, P.J., 2009. Is slow walking more stable? J. Biomech. 42, 1506–1512.
Bruijn, S.M., Van Impe, A., Duysens, J., Swinnen, S.P., 2014. White matter microstructural organization and gait stability in older adults. Front. Aging Neurosci. 6, 104.
Chamberlain, J.D., Deriaz, O., Hund-Georgiadis, M., Meier, S., Scheel-Sailer, A., Schubert, M., Stucki, G., Brinkhof, M.W., 2015. Epidemiology and contemporary risk profile of traumatic spinal cord injury in Switzerland. Inj. Epidemiol. 2, 1–11.
Chini, G., Ranavolo, A., Draicchio, F., Casali, C., Conte, C., Martino, G., Leonardi, L., Padua, L., Coppola, G., Pierelli, F., 2017. Local stability of the trunk in patients with degenerative cerebellar ataxia during walking. Cerebellum 16, 26–33.
Dingwell, J.B., 2006. Lyapunov exponents. Wiley Encyclopedia of Biomedical Engineering.
Evans, W.J., Lexell, J., 1995. Human aging, muscle mass, and fiber type composition. J. Gerontol. A Biol. Sci. Med. Sci. 50, 11–16.
Friedman, J.H., 1991. Multivariate adaptive regression splines. Ann. Stat., 1-67.
Hamacher, D., Törpel, A., Hamacher, D., Schega, L., 2016. The effect of physical exhaustion on gait stability in young and older individuals. Gait Posture 48, 137–139.
Hothorn, T., Hornik, K., Zeileis, A., 2006. Unbiased recursive partitioning: A conditional inference framework. J. Comp. Graph. Stat. 15, 651–674.
Jekabsons, G., 2016. ARESLab: Adaptive Regression Splines toolbox for Matlab/Octave available at http://www.cs.rtu.lv/jekabsons/.
Kang, H.G., Dingwell, J.B., 2008. Effects of walking speed, strength and range of motion on gait stability in healthy older adults. J. Biomech. 41, 2899–2905.
Kurz, M.J., Markopoulou, K., Stergiou, N., 2010. Attractor divergence as a metric for assessing walking balance. Nonlinear Dynamics Psychol. Life Sci. 14, 151.
Lockhart, T.E., Liu, J., 2008. Differentiating fall-prone and healthy adults using local dynamic stability. Ergonomics 51, 1860–1872.

Reynard, F., Terrier, P., 2014. Local dynamic stability of treadmill walking: Intrasession and week-to-week repeatability. J. Biomech. 47, 74–80.
Reynard, F., Terrier, P., 2015. Role of visual input in the control of dynamic balance: variability and instability of gait in treadmill walking while blindfolded. Exp. Brain Res. 233, 1031–1040.
Reynard, F., Vuadens, P., Deriaz, O., Terrier, P., 2014. Could local dynamic stability serve as an early predictor of falls in patients with moderate neurological gait disorders? A reliability and comparison study in healthy individuals and in patients with paresis of the lower extremities. PLoS One 9, e100550.
Stenum, J., Bruijn, S.M., Jensen, B.R., 2014. The effect of walking speed on local dynamic stability is sensitive to calculation methods. J. Biomech. 47, 3776–3779.
Strobl, C., Boulesteix, A.-L., Kneib, T., Augustin, T., Zeileis, A., 2008. Conditional variable importance for random forests. BMC Bioinform. 9, 307.
Strobl, C., Boulesteix, A.-L., Zeileis, A., Hothorn, T., 2007. Bias in random forest variable importance measures: Illustrations, sources and a solution. BMC Bioinform. 8, 25.
Strobl, C., Hothorn, T., Zeileis, A., 2009a. Party on! R J. 1, 14–17.
Strobl, C., Malley, J., Tutz, G., 2009b. An introduction to recursive partitioning: rationale, application, and characteristics of classification and regression trees, bagging, and random forests. Psychol. Methods 14, 323.
Talbot, L.A., Musiol, R.J., Witham, E.K., Metter, E.J., 2005. Falls in young, middle-aged and older community dwelling adults: perceived cause, environmental factors and injury. BMC Public Health 5, 86.
Terrier, P., Dériaz, O., 2013. Nonlinear dynamics of human locomotion: effects of rhythmic auditory cueing on local dynamic stability. Front. Physiol. 4.
Terrier, P., Reynard, F., 2015. Effect of age on the variability and stability of gait: a cross-sectional treadmill study in healthy individuals between 20 and 69 years of age. Gait Posture 41, 170–174.
Thompson, B., 1995. Stepwise regression and stepwise discriminant analysis need not apply here: A guidelines editorial. Educ. Psychol. Meas. 55, 525–534.
Toebes, M.J., Hoozemans, M.J., Furrer, R., Dekker, J., van Dieen, J.H., 2012. Local dynamic stability and variability of gait are associated with fall history in elderly subjects. Gait Posture 36, 527–531.
Vieira, M.F., Rodrigues, F.B., de Sa, E.S.G.S., Magnani, R.M., Lehnen, G.C., Campos, N.G., Andrade, A.O., 2017. Gait stability, variability and complexity on inclined surfaces. J. Biomech. 54, 73–79.
Wu, Y., Li, Y., Liu, A.-M., Xiao, F., Wang, Y.-Z., Hu, F., Chen, J.-L., Dai, K.-R., Gu, D.-Y., 2016. Effect of active arm swing to local dynamic stability during walking. Hum. Mov. Sci. 45, 102–109.

# Tables

**-none-**

# Figure captions

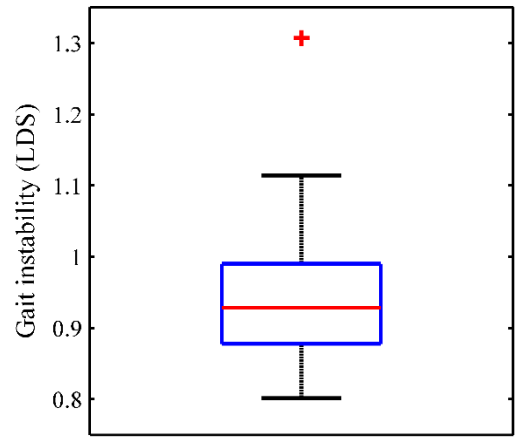

**Fig. 1.** Distribution of gait stability results measured for the 100 participants. LDS: local dynamic stability.

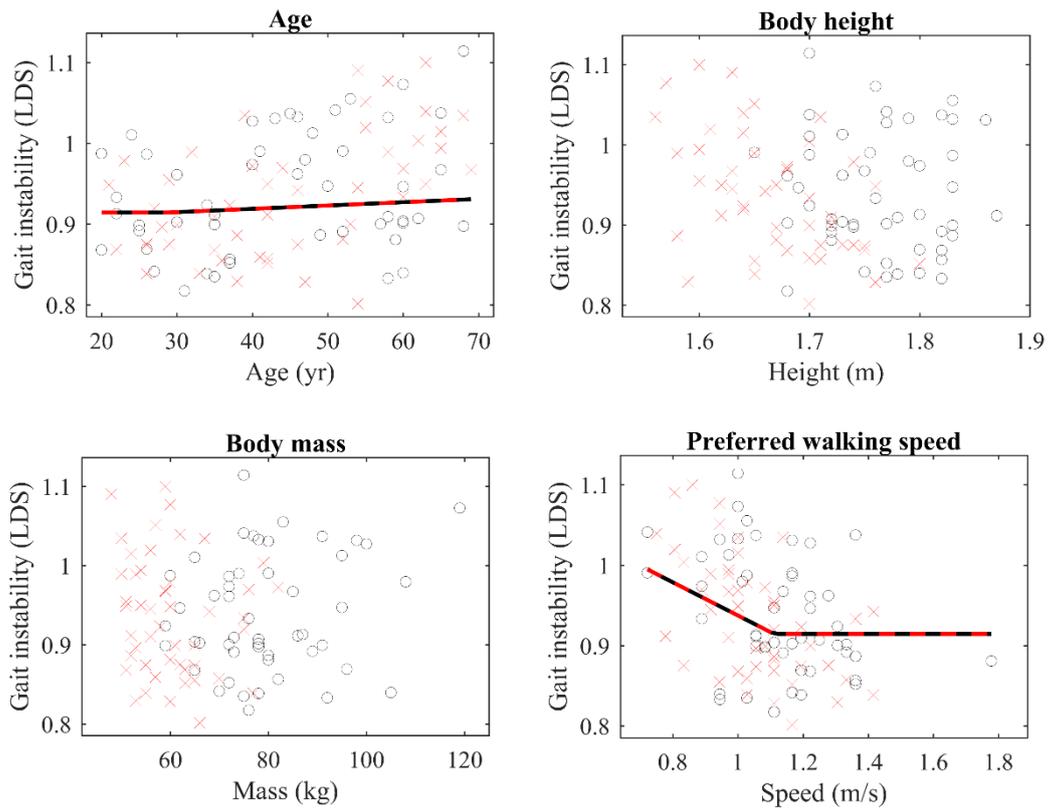

**Fig. 2.** Scatter plots and results of the multivariate adaptive regression splines (MARS) analysis. N = 99. The MARS equation is given below the plots. The lines show the output of the MARS model for relevant predictors, with others held constant. "Max" is a function that returns the largest element. Red crosses: females; black circles: males. LDS: local dynamic stability. BF1: basis function 1.

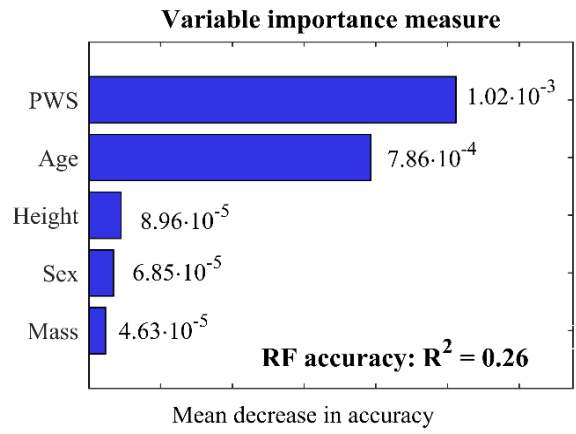

**Fig. 3.** Results of the random forest (RF) analysis. N = 99.

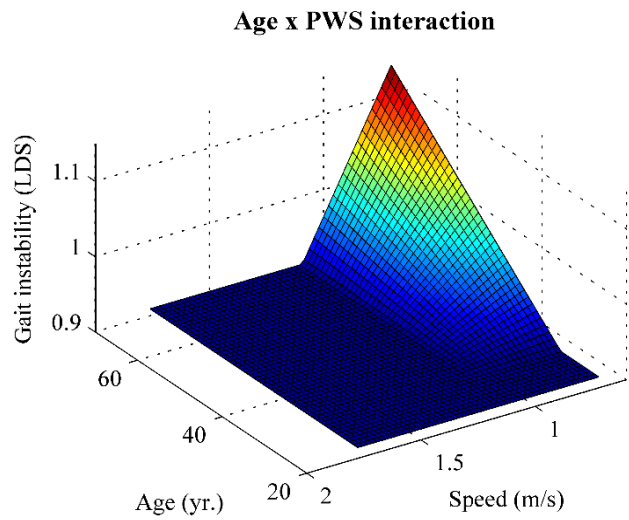

**Fig. 4** MARS results: interaction between age and preferred walking speed (PWS). LDS: Local dynamic stability. The equation is LDS=0.914+0.0148*BF1 with BF1 = *max*(0, 1.11-Speed) * *max*(0, Age-29), *max* being a function that returns the largest element.

# Determinants of gait stability while walking on a treadmill: a machine learning approach

Fabienne Reynard & Philippe Terrier

## Supplementary Analyzes

Linear regression model: quadratic terms and interactions
    Lds ~ [Linear formula with 20 terms in 5 predictors]
Estimated Coefficients:

|  | Estimate | SE | tStat | pValue |
|---|---|---|---|---|
| (Intercept) | 10.5265 | 6.8917 | 1.5274 | 0.1307 |
| Age | 0.0318 | 0.0247 | 1.2842 | 0.2028 |
| Sex_1 | -0.5709 | 0.9029 | -0.6324 | 0.529 |
| Height | -11.9666 | 7.9284 | -1.5093 | 0.1352 |
| Weight | -0.008 | 0.0358 | -0.2228 | 0.8243 |
| PWS | 1.2243 | 1.664 | 0.7358 | 0.464 |
| Age:Sex_1 | -0.0001 | 0.0014 | -0.0505 | 0.9598 |
| Age:Height | -0.0195 | 0.014 | -1.389 | 0.1687 |
| Age:Weight | 0.0001 | 0.0001 | 1.1492 | 0.2539 |
| Age:PWS | -0.0012 | 0.0035 | -0.3337 | 0.7395 |
| Sex_1:Height | 0.1359 | 0.4931 | 0.2756 | 0.7836 |
| Sex_1:Weight | 0.0034 | 0.0028 | 1.205 | 0.2318 |
| Sex_1:PWS | 0.0562 | 0.1148 | 0.4893 | 0.626 |
| Height:Weight | -0.0132 | 0.0207 | -0.6363 | 0.5264 |
| Height:PWS | -1.2913 | 0.9803 | -1.3173 | 0.1916 |
| Weight:PWS | 0.0146 | 0.0061 | 2.3942 | 0.019 |
| Age^2 | 0 | 0 | -0.2057 | 0.8375 |
| Height^2 | 4.3159 | 2.4531 | 1.7594 | 0.0824 |
| Weight^2 | 0.0001 | 0.0001 | 1.38 | 0.1715 |
| PWS^2 | -0.0836 | 0.1466 | -0.5704 | 0.57 |

Number of observations: 99, Error degrees of freedom: 79
Root Mean Squared Error: 0.061
R-squared: 0.451,  Adjusted R-Squared: 0.318
F-statistic vs. constant model: 3.41, p-value = 6.51e-05

Linear regression model: interactions

    Lds ~ [Linear formula with 16 terms in 5 predictors]

Estimated Coefficients:

|               | Estimate | SE     | tStat   | pValue |
|---------------|----------|--------|---------|--------|
| (Intercept)   | 0.422    | 3.0773 | 0.1371  | 0.8913 |
| Age           | 0.0388   | 0.0179 | 2.1697  | 0.0329 |
| Sex_1         | 0.0362   | 0.8121 | 0.0446  | 0.9645 |
| Height        | 1.0426   | 1.7749 | 0.5874  | 0.5585 |
| Weight        | -0.0344  | 0.0314 | -1.0964 | 0.2761 |
| PWS           | 0.4454   | 1.5183 | 0.2933  | 0.77   |
| Age:Sex_1     | -0.0002  | 0.0013 | -0.1527 | 0.879  |
| Age:Height    | -0.0251  | 0.0109 | -2.2947 | 0.0243 |
| Age:Weight    | 0.0001   | 0.0001 | 2.0215  | 0.0465 |
| Age:PWS       | -0.0017  | 0.0034 | -0.4914 | 0.6244 |
| Sex_1:Height  | -0.1805  | 0.4201 | -0.4296 | 0.6686 |
| Sex_1:Weight  | 0.0023   | 0.0023 | 0.99    | 0.3251 |
| Sex_1:PWS     | 0.0788   | 0.1129 | 0.6982  | 0.487  |
| Height:Weight | 0.0101   | 0.0169 | 0.5994  | 0.5505 |
| Height:PWS    | -0.7874  | 0.9033 | -0.8717 | 0.3859 |
| Weight:PWS    | 0.011    | 0.0053 | 2.0697  | 0.0416 |

Number of observations: 99, Error degrees of freedom: 83
Root Mean Squared Error: 0.0612
R-squared: 0.42,  Adjusted R-Squared 0.315
F-statistic vs. constant model: 4, p-value = 2.11e-05

Stepwise backward multiple regression using Bayes Information Criteria (BIC)

1. Removing Age:Sex, BIC = -207.78
2. Removing Age^2, BIC = -212.32
3. Removing Sex:Height, BIC = -216.83
4. Removing Age:PWS, BIC = -221.34
5. Removing Sex:PWS, BIC = -225.6
6. Removing PWS^2, BIC = -229.76
7. Removing Height:Weight, BIC = -233.32
8. Removing Weight^2, BIC = -235.82
9. Removing Sex:Weight, BIC = -238.89
10. Removing Height:PWS, BIC = -241.53
11. Removing Weight:PWS, BIC = -243.36
12. Removing Age:Weight, BIC = -246.23
13. Removing Weight, BIC = -249.41
14. Removing Age:Height, BIC = -252.12
15. Removing Height^2, BIC = -253.2
16. Removing Height, BIC = -254.21
17. Removing Sex, BIC = -256.63

Linear regression model:
  Lds ~ 1 + Age + PWS

|  | Estimate | SE | tStat | pValue |
|---|---|---|---|---|
| (Intercept) | 1.0337 | 0.0466 | 22.1884 | 1.43E-39 |
| Age | 0.0018 | 0.0004 | 4.0075 | 0.00012121 |
| PWS | -0.1612 | 0.0363 | -4.4382 | 2.42E-05 |

Number of observations: 99, Error degrees of freedom: 96
Root Mean Squared Error: 0.0627
R-squared: 0.295, Adjusted R-Squared: 0.28
F-statistic vs. constant model: 20.1, p-value = 5.15e-08